# Generalized Parity-Time Symmetry Condition for Enhanced Sensor Telemetry

Pai-Yen Chen[1,*], Maryam Sakhdari[1], Mehdi Hajizadegan[1], Qingsong Cui[1], Mark Cheng[1], Ramy El-Ganainy[2,3], and Andrea Alù[4,5*]

[1]Department of Electrical and Computer Engineering, Wayne State University, Detroit, MI 48202, USA

[2]Department of Physics, Michigan Technological University, Houghton, MI 49931, USA

[3]Department of Electrical and Computer Engineering, Michigan Technological University, Houghton, MI 49931, USA

[4]Advanced Science Research Center, City University of New York, New York, NY 10031, USA

[5]Graduate Center of the City University of New York, New York, NY 10016, USA

**Wireless sensors based on micromachined tunable resonators are important in a variety of applications, ranging from medical diagnosis to industrial and environmental monitoring. The sensitivity of these devices is, however, often limited by their low quality ($Q$) factor. Here, we introduce the concept of isospectral party–time–reciprocal scaling (*PTX*) symmetry and show that it can be used to build a new family of radiofrequency wireless microsensors exhibiting ultrasensitive responses and ultrahigh resolution, which are well beyond the limitations of conventional passive sensors. We show theoretically, and demonstrate experimentally using microelectromechanical-based wireless pressure sensors, that *PTX*-symmetric electronic systems share the same eigenfrequencies as their parity–time (*PT*)-symmetric counterparts, but crucially have different circuit profiles and eigenmodes. This simplifies the electronic circuit design and enables further enhancements to the extrinsic $Q$-factor of the sensors.**

The wireless monitoring of physical, chemical and biological quantities is essential in a range of medical and industrial applications in which physical access and wired connections would introduce significant limitations. Examples include sensors that are required to operate in harsh environments, and those that are embedded in, or operate in the vicinity of, human bodies [1]. Telemetric sensing based on compact, battery-less wireless sensors is one of the most feasible ways to perform contactless continuous measurements in such applications. The first compact passive wireless sensor was proposed in1967 [2], and used a miniature spiral inductor ($L$) and a pressure-sensitive capacitor ($C$) to build a resonant sensor that could measure the fluid pressure inside the eye (an intraocular pressure sensor). The idea was based on a mechanically adjusted capacitor (or varactor), which has been an effective way of tuning resonant circuits since the advent of the radio [3]. Despite this, wireless capacitive sensing technology has experienced a rapid expansion only in the last two decades, due to the development of microelectromechanical systems (MEMS), nanotechnology and wireless technology [4]-[8].

Recently, low-profile wireless sensors based on passive *LC* oscillating circuitry (typically a series *RLC* tank) have been used to measure pressure [5],[6], strain [7], drug delivery [8], temperature, and chemical reactions [1]. The working principle of these passive *LC* sensors is typically based on detecting concomitant resonance frequency shifts, where the quantity to be measured detunes capacitive or inductive elements of the sensor. This could occur, for example, through mechanical deflections of electrodes, or variations of the dielectric constant. In general, the readout of wireless sensors relies on mutual inductive coupling (Fig. 1a), and the sensor information is encoded in the reflection coefficient. Such telemetric sensor systems can be modelled using a simple equivalent circuit model, in which the compact sensor is represented by a series resonant *RLC* tank, where the resistance *R* takes into account the power dissipation of the sensor (Fig. 1a).

Although there has been continuous progress in micro- and nano-machined sensors in recent years, the basics of the telemetric readout technique remain essentially unchanged since its invention. Nonetheless, improving the detection limit is often hindered by the available levels of *Q*-factor, the sensing resolution and the sensitivity related to the spectral shift of resonance in response to variations of the physical property to be measured. In particular, modern *LC* microsensors based on thin-film resonators or actuators usually have a low modal *Q*-factor, due to relevant power dissipations caused by skin effects, Eddy currents and the electrically lossy surrounding environment (such as biological tissues) [9]. A sharp, narrowband reflection dip has been a long-sought goal for inductive sensor telemetry, because it could lead to superior detection and great robustness to noises.

In this Article, we introduce a generalized parity-time (*PT*)-symmetric telemetric sensing technique, which enables new mechanisms to manipulate radio frequency (RF) interrogation between the sensor and the reader, to boost the effective *Q*-factor and sensitivity of wireless microsensors. We implement this sensing technique using MEMS-based wireless pressure sensors operating in the RF spectrum.

**Generalized PT-Symmetry**

The concept of *PT*-symmetry was first proposed by Bender in the context of quantum mechanics [10] and has been extended to classical wave systems, such as optics [11]-[13], owing to the mathematical isomorphism between Schrodinger and Helmholtz wave equations. *PT*-symmetric optical structures with balanced gain and loss have unveiled several exotic properties and applications, including unidirectional scattering [14],[15], coherent perfect absorber-laser [16],[17], single-mode micro-ring laser [18]-[20], and optical non-reciprocity [21][24]. Inspired by optical schemes, other *PT*-symmetric systems in electronics (sub-radiofrequency, 30 KHz and below [25]-[27]), acoustics [28] and optomechanics [29],[30] have also been recently reported. The exceptional points arising in these systems, found at the bifurcations of eigenfrequencies near the *PT*-phase transition, show potential to enhance the sensitivity of photonic sensors [31]-[35].

In principle, exceptional points and bifurcation properties of a *PT*-symmetric system can be utilized also to enhance sensor telemetry, represented by the equivalent circuit in Fig. 1b with $x = 1$. In this case, the *PT*-symmetry condition is achieved when the gain and loss parameters, namely $-R$ and $R$, are delicately balanced, and the reactive components, $L$ and $C$, satisfy mirror symmetry: that is, the impedances of the active and passive circuit tanks, multiplied by $i$, are complex conjugates of each other at the frequency of interest. Similar to earlier experiments in optical systems [22], the realization of *PT*-symmetry in a telemetric sensor system is expected to exhibit real eigenfrequencies in the exact symmetry phase. This leads to sharp and deep resonances, beyond the limitations discussed above for passive systems, thus providing improved spectral resolution and modulation depth for sensing. Despite this advantage of traditional *PT*-symmetric systems, practical implementations for the sensor telemetry may encounter difficulties in achieving an exact conjugate impedance profile. For instance, given the limited area of medical bioimplants and MEMS-based sensors, the inductance of the sensor's microcoil $L_S$ is usually smaller than the one of the reader's coil $L_R$. Although downscaling the reader coil can match $L_R$ to $L_S$, this would reduce the mutually inductive coupling and degrade the operation of the wireless sensor. Therefore, it is highly desirable to have extra degrees of freedom that allow arbitrary scaling of the coil inductance and other parameters (for example, capacitance and equivalent negative resistance) in the reader, to optimize the wireless interrogation and facilitate the electronic circuit integration.

To overcome these difficulties, and at the same time significantly improve the sensing capabilities of telemetric sensors, we also introduce here the idea of *PTX*-symmetric telemetry (Fig. 1b). This *PTX*-symmetric electronic system consists of an active reader (equivalently, a –*RLC* tank), wirelessly interrogating a passive microsensor (*RLC* tank) via the inductive coupling. Here, the equivalent series –*R* is achieved with a Colpitts-type circuit (Fig. 1b), which acts as a negative resistance converter (NRC) (see Supplementary Note 1 for detailed design, analysis and characterization of the circuit). By suitably scaling the values of –*R*, *L* and *C* in the active reader, the system can be made invariant under the combined parity transformation $\mathcal{P}$ ($q_1 \leftrightarrow q_2$), time-reversal transformation $\mathcal{T}$ ($t \rightarrow -t$), and reciprocal scaling $\mathcal{X}$ ($q_1 \rightarrow x^{1/2} q_1$, $q_2 \rightarrow x^{-1/2} q_2$), where $q_1$ ($q_2$) corresponds to the charge stored in the capacitor in the –*RLC* (*RLC*) tank, and $x$ is the reciprocal-scaling coefficient, an arbitrary positive real number. In the following analysis, we will prove that the introduced $\mathcal{X}$ transformation allows the operation of a system with unequal gain and loss coefficients (also an asymmetric reactance distribution), while exhibiting an eigenspectrum that is identical to the one of the *PT*-symmetric system. Crucially, the scaling operation $\mathcal{X}$ offers an additional degree of freedom in sensor and reader designs, overcoming the mentioned space limitations of microsensors that pose challenges in realizing *PT*-symmetric telemetry. Even more importantly, while the scaling provided by the $\mathcal{X}$ operator leaves the eigenspectrum unchanged, it leads to linewidth sharpening and thus boosts the extrinsic *Q*-factor, the sensing resolution, and the overall sensitivity.

As we demonstrate below, the effective Hamiltonians of *PTX* and *PT* systems are related by a mathematical similarity transformation. We start by considering Kirchoff's law of the equivalent circuit representation of the *PTX* telemetric sensor system (Fig. 1b) cast in the form of Liouville-type equation $\partial_\tau \Psi = \mathcal{L}\Psi$ (ref. [25]) governing the dynamics of this coupled *RLC/–RLC* dimer, where the Liouvillian $\mathcal{L}$ is given by

$$\mathcal{L} = \begin{pmatrix} 0 & 0 & 1 & 0 \\ 0 & 0 & 0 & 1 \\ -\dfrac{1}{1-\kappa^2} & \dfrac{1}{\sqrt{x}}\dfrac{\kappa}{1-\kappa^2} & \dfrac{1}{\gamma(1-\kappa^2)} & \dfrac{1}{\sqrt{x}}\dfrac{\kappa}{\gamma(1-\kappa^2)} \\ \sqrt{x}\dfrac{\kappa}{1-\kappa^2} & -\dfrac{1}{1-\kappa^2} & -\sqrt{x}\dfrac{\kappa}{\gamma(1-\kappa^2)} & -\dfrac{1}{\gamma(1-\kappa^2)} \end{pmatrix}, \quad (1)$$

and $\Psi \equiv (q_1, q_2, \dot{q}_1, \dot{q}_2)^T$, $\tau \equiv \omega_0 t$, the natural frequency of an isolated lossless *LC* tank $\omega_0 = 1/\sqrt{LC}$, the coupling strength between the active and passive tanks $\kappa = M/\sqrt{L_R L_S}$, $L_R = xL$, $L_S = L$, and the dimensionless non-Hermiticity parameter $\gamma = R^{-1}\sqrt{L/C} = (x/R')^{-1}\sqrt{(xL)/(C/x)}$; here all frequencies are measured in units of $\omega_0$. The active and passive tanks have the same non-Hermiticity parameter $\gamma$, regardless of the value of $x$ (*PT* or *PTX* system). From equation (1), we can define an effective Hamiltonian $H = i\mathcal{L}$ with non-Hermitian form (that is, $H^\dagger \neq H$). Such a non-Hermitian Hamiltonian system is invariant under a combined $\mathcal{PTX}$ transformation, with

$$\mathcal{P} = \begin{pmatrix} \sigma_x & 0 \\ 0 & \sigma_x \end{pmatrix}, \quad (2a)$$

$$\mathcal{T} = \begin{pmatrix} \mathbf{1} & 0 \\ 0 & -\mathbf{1} \end{pmatrix}\mathcal{K}, \quad (2b)$$

$$\mathcal{X} = \mathbf{1} \otimes x_0 \text{ and } x_0 = \begin{pmatrix} x^{1/2} & 0 \\ 0 & x^{-1/2} \end{pmatrix}, \quad (2c)$$

where $\sigma_x$ is the Pauli matrix, $\mathbf{1}$ is the identity matrix, $\mathcal{K}$ performs the operation of complex conjugation, and $(\mathcal{PTX})^2 = \mathbf{1}$. The Hamiltonian and eigenmodes of the *PTX* system are related to those of the *PT* system $(H', \Psi')$ through the similarity transformation $H = S^{-1} H' S$ and $\Psi = S^{-1}\Psi'$, where $S$ is an invertible 4-by-4 matrix $S = \mathbf{1} \otimes \zeta$ and $\zeta = \begin{pmatrix} x^{1/2} & 0 \\ 0 & 1 \end{pmatrix}$. As a result, *PTX* and *PT* systems share the same eigenfrequencies, but possess different eigenmodes. Moreover, $H$ commutes with the transformed operators $\breve{\mathcal{P}} = S^{-1}\mathcal{P}S$ and $\breve{\mathcal{T}} = S^{-1}\mathcal{T}S = \mathcal{T}$, i.e., $\left[\breve{\mathcal{P}}\breve{\mathcal{T}}, H\right] = 0$, where $\breve{\mathcal{P}}$ performs the combined operations of parity and reciprocal scaling: $x^{1/2} q_1 \leftrightarrow x^{-1/2} q_2$. After some mathematical manipulations, we obtain $\breve{\mathcal{P}}\breve{\mathcal{T}} = \mathcal{PTX}$, and, therefore, $H$ commutes also with $\mathcal{PTX}$ (that is $\left[\mathcal{PTX}, H\right] = 0$). In the limit when the scaling coefficient $x = 1$, the *PTX*-symmetric system converges into the traditional *PT*-symmetric system. Hence, the *PTX*-symmetry can be regarded as a generalized group of the *PT*-symmetry.

### *PT/PTX*-symmetric telemetric microsensor systems

We designed and realized the sensor using a micromachined parallel-plate varactor connected in series to a micromachined planar spiral inductor and also a parasitic resistance (Fig. 1b). Figure 2a-c shows schematic diagrams and a photograph of the realized device, together with its detailed surface profiles characterized by scanning white-light interferometry (SWLI) (see Supplementary Notes 1 and 2 for design and fabrication details). The sensor was encapsulated with epoxy polyamides and connected to an air compressor, and a microprocessor-controlled regulator was used to vary the internal pressure inside the MEMS microcavity from 0mmHg to 200mmHg. This procedure simulates, for instance, pressure variations inside the human eye [6] (see Methods for the detailed measurement set-up). The sensor can be seen as a tunable passive *RLC* tank, in which the applied pressure mechanically deforms the floating electrode of the varactor (Fig. 2a), causing a change in the total capacitance. Figure 2d presents the extracted capacitance as a function of the internal pressure, with insets showing the corresponding cross-sectional SWLI images (see Supplementary Note 1 for the extraction of *RLC* values). The measurement results agree well with theoretical predictions, revealing that the capacitance is reduced by increasing the applied pressure.

In our first set of experiments, we designed an active reader, which, together with the passive microsensor, forms the *PT*-symmetric dimer circuit. We investigate the evolution of complex eigenfrequencies and reflection spectra as we vary $\gamma$ and $\kappa$. In our measurements, the sensor was fixed on an *XYZ* linear translation stage used to precisely control $\kappa$. For a specific value of $\kappa$, $\gamma$ was tuned by the equivalent capacitance of the microsensor, responsible for the applied pressure. On the reader side, the voltage-controlled impedance converter provides an equivalent negative resistance, whose magnitude is set equal to $-(R-Z_0)$, where the sensor's effective resistance $R$ was measured to be ~150 Ω and $Z_0$ is the source impedance of the RF signal generator (for example, vector network analyzer (VNA) used in the experiment, with $Z_0 = 50\,\Omega$) connected in series to the active reader. We note that, in the closed-loop analysis, an external RF source can be modelled as a negative resistance $-Z_0$, as it supplies energy to the system [3]. When the sensor's capacitance changed, the voltage-controlled varactor in the reader circuit was adjusted accordingly to maintain the *PT*-symmetry condition (see Supplementary Note 1 for details of reader design). Wireless pressure sensing was performed by monitoring in situ the shift of resonance in the reflection spectrum across 100–350 MHz. In our measurements, a clear eigenfrequency bifurcation with respect to $\gamma$ and $\kappa$ of the *PT*-symmetric system was observed (as shown in Fig. 3a) and the agreement between experimental results (dots) and theory (colored contours) is excellent; a detailed theoretical analysis of the critical points is provided in the Methods. At the exceptional point $\gamma_{EP}$, real eigenfrequencies branch out into the complex plane. In the region of interest $\gamma \in [\gamma_{EP}, \infty]$, the eigenfrequencies are purely real ($\omega \in \mathbb{R}$) (Fig. 3a) and $\mathcal{PT}\Psi' = \Psi'$, such that the PT-symmetry condition is exactly met in the so-called exact PT-symmetric phase. In this phase, the oscillation occurs at two distinct eigenfrequencies corresponding to sharp reflection dips (Fig. 3c). Before passing $\gamma_{EP}$, the system is in its broken PT-symmetric phase, where complex eigenfrequencies ($\omega \in \mathbb{C}$) exist in the form of complex conjugate pairs, and the PT-symmetry of eigenmodes is broken, namely $\mathcal{PT}\Psi' \neq \Psi'$. The system exhibits a phase transition when the non-Hermiticity parameter exceeds the critical value $\gamma_{EP}$, at which point the non-Hermitian degeneracy can unveil several counterintuitive features, such as the unidirectional reflectionless transparency [14],[28] and the singularity-enhanced sensing [31]-[35].

To better illustrate the system response, we plot the measured reflection spectra, where $\gamma$ is fixed to 2.26 (corresponding to an applied pressure of 100 mmHg), while $\kappa$ is continuously varied from 0.4 to 0.5 (Fig. 3c). The evolution of the resonant response clearly identifies the eigenfrequency transition (Fig. 3a). In the weak coupling region, the system operates in the broken PT-symmetric phase, quantified by $\kappa < \kappa_{PT}$, and its complex eigenfrequency results in a weak and broad resonance. This can be explained by the fact that, if the coupling strength is weak, the energy in the active –RLC tank cannot flow fast enough into the passive RLC tank to compensate for the absorption, thereby resulting in a non-equilibrium system with complex eigenfrequencies. If the coupling strength exceeds a certain threshold, the system can reach equilibrium, since the energy in the active tank can flow fast enough into the passive one to compensate its power dissipation. From Fig. 3a, we observe that at higher $\kappa$, the threshold of $\gamma$ for the phase transition ($\gamma_{EP}$) can be reduced. As a result, a *PT*-symmetric telemetric sensor system, if designed properly to work in the exact *PT*-symmetric phase quantified by $\kappa > \kappa_{PT}$, can exhibit sharp and deep resonant reflection dips, ensuring high sensitivity with electrical noise immunity. From the circuit viewpoint, the reflectionless property in the one-port measurement is due to impedance matching. In the exact *PT*-symmetric phase with real eigenfrequencies, the input impedance looking into the active reader can be matched to the generator impedance $Z_0$ at the eigenfrequencies (or resonance frequencies), leading to the dips observed in the reflection spectrum.

We also note that the splitting of the Riemann surface outlined in Fig. 3a may lead to an interesting topological response, implying a dramatic shift of the resonance frequency when $\gamma$ is altered by pressure-induced capacitance changes in the microsensor ($\gamma \propto C^{-1/2}$). It is interesting to compare these results with those obtained with a conventional fully passive telemetric sensing scheme (Fig. 3b) [4]-[9], where the negative-resistance converter and varactors are removed from the active reader, leaving a coil antenna to interrogate the same pressure sensor. In this case, the eigenfrequency of the conventional passive system is always complex (Fig. 3b), no matter how $\gamma$ and $\kappa$ are varied, as expected for a lossy resonator, and the eigenfrequency surface is rather flat for both real and imaginary parts when compared with the PT-symmetric system (Fig. 3a). Figure 4a,b presents the evolution of the reflection spectra for the two sensing systems; here $\kappa$ is fixed to 0.5 and $\gamma$ is varied by changing the applied pressure (20, 40, 70, and 100 mmHg). The bifurcation of eigenfrequency in the *PT*-symmetric system (Fig. 4b) leads to the formation of two eigenmodes with sharp reflection dips, whose spectral shifts in response to $\gamma$ can be dramatic and coincides with the topological phase transition shown in Fig. 3a. On the other hand, the passive system (Fig.

4a) exhibits a broad resonance, associated with a low sensing resolution, and a less observable change in the resonance frequency. It is evident that a *PT*-symmetric telemetric sensor can provide largely superior sensitivity when compared with conventional passive ones [4][9], as it achieves not only a finer spectral resolution in light of a higher *Q*-factor, but also more sensitive frequency responses (Fig. 5a).

Next, we explore the functionality of the *PTX*-symmetric sensor within the same telemetry platform. Unlike the *PT*-symmetric system, the reciprocal scaling in the *PTX* system breaks the mirror symmetry of the effective $|\pm R|$, *L*, and *C*; namely, their values in the sensor and the reader can be quite different for large or small values of *x*. In our experiments, the same MEMS-based pressure sensor was now paired with a new type of reader (Fig. 1b), whose equivalent circuit is similar to the reader used in Fig. 4b, but with all elements scaled following the rule: $-R \to -xR$, $L \to xL$, and $C \to x^{-1}C$. This realizes a *PTX*-symmetric telemetry system that has a non-Hermitian Hamiltonian *H* (equation (1)) commuting with $\mathcal{PTX}$ (equation (2)). We have tested different values of *x* to investigate its effect on eigenfrequencies; here $\kappa$ was fixed to 0.5 in different setups. Figure 5a shows the real and imaginary parts of eigenfrequencies against $\gamma$ for *PTX*-symmetric telemetric sensor systems with $x = 3$, 1/3, and 1. We note that $x = 1$ corresponds to the *PT*-symmetric system discussed before.

We observe that a non-Hermitian *PTX*-symmetric Hamiltonian also supports real eigenfrequencies in the exact *PTX*-symmetric phase, thus leading to sharp and deep resonant reflection dips. As discussed earlier, in spite of the introduction of the $\mathcal{X}$ operator, the *PTX*-symmetric system and its *PT*-symmetric counterpart possess exactly the same eigenspectrum and bifurcation points, as clearly seen in Fig. 5a. In the *PTX* system, there is also a clear transition between the exact *PTX*-symmetric phase ($\mathcal{PTX}\Psi = \Psi$) and the broken *PTX*-symmetric phase ($\mathcal{PTX}\Psi \neq \Psi$), which are respectively characterized by real and complex eigenfrequencies. The theoretical and experimental results in Fig. 5a imply that the spectral shift of resonance associated with the exceptional-point singularity in a *PT*-symmetric sensor can be likewise obtained in a *PTX*-symmetric sensor, as the same eigenspectrum is shared. We note that the *PT* and *PTX* systems, although sharing the same eigenspectrum, can have different eigenmodes; that is, $\Psi = S^{-1}\Psi'$ and $S$ is correlated with *x*. Figure 5b presents reflection spectra for the *PTX*-symmetric telemetric sensor with *x*=3, under different applied pressures. Due to the scaling operation $\mathcal{X}$ in the *PTX*-symmetric system, it is possible to further reduce the linewidth of the reflection dip and achieve a finer sensing resolution by increasing the value of *x*. In contrast to the case $x > 1$, $x < 1$ results in broadening of the resonance linewidth and thus a lowered *Q*-factor. We note that the input impedance (looking into the active reader) of *PT*- and *PTX*-symmetric telemetry systems can be identical and matched to the generator impedance $Z_0$ at their shared resonance frequencies, corresponding to reflectionless points (see Supplementary Note 3). As the frequency is away from the resonance frequency, the input impedance and reflection coefficient of *PT*- and *PTX*-symmetric systems may be very different, leading to a different resonance linewidth as a function of *x*. As a result, the *PTX*-symmetric telemetric sensor system (Fig. 5b), when compared with the PT-symmetric one (Fig. 4b), not only offers more design flexibility by removing certain physical constraints (for example, mirror-symmetric |±*R*|, *L* and *C* in the mutually coupled circuit), but also could support greater resolution, sensitivity and potentially longer interrogation distance enabled by the optimally designed self and mutual inductances of coils. Most importantly, both systems exhibit the same eigenspectrum and exceptional point. Ideally, in the exact *PTX*-symmetry phase, there is no fundamental limit to the *Q*-factor enhancement. In the extreme case when *x* approaches infinity, the resonance linewidth becomes infinitesimally narrow, namely the *Q*-factor is close to infinity, provided that such a reader circuit can be realized. However, in reality, the –R, L and C values of electronic devices have their own limits.

For generality, a microsensor (negative-resistance converter) can in principle be decomposed into a series or parallel equivalent *RLC* (*–RLC*) tank, and either choice is formally arbitrary, depending on the sensor and circuit architectures and on the kind of excitation (that is, impressed voltage or current source). The concept of *PTX*-symmetry can also be generalized to an electronic dimer utilizing the parallel circuit configuration, whose *PT*-symmetric counterpart has been demonstrated [25],[26]. It may also be possible to enhance the performance and resolution of a wireless resonant sensor modelled by a parallel *RLC* tank if the sensor is interrogated by a parallel *–RLC* tank [25],[26], to satisfy the *PTX*-symmetry condition (see Supplementary Note 3 for an example of the *PTX*-symmetric parallel circuit).

It is important to note that, in the exact symmetry phase of the *PTX*-symmetric system, although the gain and loss parameters (*–xR* and *R*) are not equal, the net power gained in the active tank and the one dissipated in the passive tank are balanced, similar to the *PT*-symmetric case. In the closed-loop analysis, the power loss in the passive tank $P_{loss} = |\dot{q}_2|^2 R/2$, while the power gained in the active tank $P_{gain} = |\dot{q}_1|^2 (xR - Z_0)/2 + |\dot{q}_1|^2 Z_0/2$ (where the first term accounts for power gained from the negative-resistance device and the second term corresponds to the external energy source modelled as a negative resistance $-Z_0$). Since the *PTX*-symmetry enforces the condition $\dot{q}_1 = \dot{q}_2/\sqrt{x}$, gain and dissipation are always balanced in this system (that is, $P_{gain} = P_{loss}$), regardless of the value of *x*. Therefore, although this generalized *PT*-symmetric system allows for arbitrary scaling of the gain and loss parameters (*–R* and *R* here), the gain–loss power balance is maintained in the exact symmetry condition, as expected by the fact that the eigenvalues are real. However, greater design flexibility on the linewidth of the response could be enabled.

Finally, it is interesting to note that in the *PTX*-symmetric system, if $x$ is sufficiently small such that $xR - Z_0 \leq 0$, both the reader and sensor circuits can be fully passive; namely, an inductively coupled *RLC/RLC* dimer is used. Such an observation is in stark contrast with what one would expect in conventional *PT*-symmetric systems, where pertinent gain or amplification is necessary to enable the associated peculiar phenomena. Figure 5c presents reflection spectra for the *PTX*-symmetric telemetric sensor system with *x*=1/3; in this case, the reader is also a passive *RLC* tank without the need of a negative-resistance or amplification device. We observe a broad resonance, as the linewidth of the reflection dip is widened by decreasing the value of *x*. This operating regime (*x*=1/3), although not necessarily of interest for enhanced sensing capabilities, provides an interesting platform to study the dynamics of exceptional points and non-Hermitian physics in a loss–loss dimer, without the need for any active component. The presented *PTX*-symmetric dimer structure may also be extended to other frequencies, including light and ultrasonic waves. For instance, one potential application of our proposed reciprocally scaling operation is to provide an additional knob to tailor the threshold gain of *PT*-symmetric single-mode lasers [18]-[20] or coherent perfect absorber-lasers [16],[17] by breaking the exact balance of gain and loss coefficients, while preserving the spectrum of eigenvalues.

**Conclusions**

We have applied *PT*-symmetry and the generalized *PTX*-symmetry introduced here to RF sensor telemetry, with a particular focus on compact wireless micro-mechatronic sensors and actuators. Our approach overcomes the long-standing challenge of implementing a miniature wireless microsensor with high spectral resolution and high sensitivity, and opens opportunities to develop loss-immune high-performance sensors, due to gain–loss interactions via inductive coupling and eigenfrequency bifurcation resulting from the *PT* (*PTX*)-symmetry. Our findings also provide alternative schemes and techniques to reverse the effects of loss and enhance the *Q*-factor of various RF systems. Through our study of *PTX*-symmetry, we have shown that even asymmetric profiles of gain and loss coefficients can yield exotic non-Hermitian physics observed in *PT*-symmetric structures. Importantly, compared to *PT*-symmetry, *PTX*-symmetry offers greater design flexibility in manipulating resonance linewidths and *Q*-factors, while exhibiting eigenfrequencies identical to the associated *PT*-symmetric system.

**Methods**

**Exceptional point and phase transitions.** Applying Kirchhoff's laws to the *PTX*-symmetric circuit in Fig. 1b leads to the following set of equations:

$$\frac{d^2 q_1}{d\tau^2} = -\frac{1}{1-\kappa^2} q_1 + \frac{1}{\sqrt{x}} \frac{\kappa}{1-\kappa^2} q_2 + \frac{1}{\gamma(1-\kappa^2)} \dot{q}_1 + \frac{1}{\sqrt{x}} \frac{\kappa}{\gamma(1-\kappa^2)} \dot{q}_2, \quad (3a)$$

$$\frac{d^2 q_2}{d\tau^2} = \sqrt{x} \frac{\kappa}{1-\kappa^2} q_1 - \frac{1}{1-\kappa^2} q_2 - \sqrt{x} \frac{\kappa}{\gamma(1-\kappa^2)} \dot{q}_1 - \frac{1}{\gamma(1-\kappa^2)} \dot{q}_2, \quad (3b)$$

which leads to the Liouvillian formalism in equation (1). After the substitution of time-harmonic charge distributions $q_n = A_n e^{i\omega\tau}$, eigenfrequencies and normal modes for this *PTX*-symmetric electronic circuit can be computed from the eigenvalue equation $(H - \omega_k \mathbf{I})\Psi_k = 0$, with $k = 1, 2, 3, 4$. The eigenfrequencies associated with the non-Hermiticity parameter $\gamma$ and coupling strength $\kappa$ can be derived as:

$$\omega_{1,2,3,4} = \pm\sqrt{\frac{2\gamma^2 - 1 \pm \sqrt{1 - 4\gamma^2 + 4\gamma^4 k^2}}{2\gamma^2(1-k^2)}} \quad (4)$$

There is a redundancy in equation (4) because positive and negative eigenfrequencies of equal magnitude are essentially identical. Equation (4) is also valid for the *PT*-symmetric system, as the eigenfrequencies in equation (4) are found to be independent of *x*. We note that if *x*=1, the *PTX*-symmetric system would degenerate into the *PT* one. The eigenmodes of the *PT*-symmetric system ($\Psi'_k$) and the *PTX*-symmetric system ($\Psi_k$) can be written as:

$$\Psi'_k = c_k \left( e^{-i\phi'_k}, e^{i\phi'_k}, -i\omega_k e^{-i\phi'_k}, -i\omega_k e^{i\phi'_k} \right)^T, \quad c_k \in \mathbb{R}; \tag{5a}$$

$$e^{2i\phi'_k} = -\frac{\gamma\left[(1-\kappa^2)\omega_k^2 - 1\right] + i\omega_k}{\kappa(\gamma + i\omega_k)}; \tag{5b}$$

$$\Psi_k = S^{-1}\Psi'_k. \tag{5c}$$

Complex eigenfrequencies would evolve with γ, unveiling three distinct regimes of behavior. The eigenfrequencies undergo a bifurcation process and branch out into the complex plane at the exceptional point (or spontaneous *PTX*-symmetry breaking point):

$$\gamma_{EP} = \frac{1}{\kappa}\sqrt{\frac{1+\sqrt{1-\kappa^2}}{2}}. \tag{6}$$

In the parametric region of interest $\gamma \in [\gamma_{EP}, \infty]$, *PTX*-symmetry is exact, rendering real eigenfrequencies and $\mathcal{PTX}\Psi_k = \Psi_k$. The region $\gamma \in [\gamma_c, \gamma_{EP}]$ is known as the broken *PTX*-symmetric phase with complex eigenfrequencies. Another crossing between the pairs of degenerate frequencies (and another branching) occurs at the lower critical point:

$$\gamma_c = \frac{1}{\kappa}\sqrt{\frac{1-\sqrt{1-\kappa^2}}{2}}. \tag{7}$$

In the sub-critical region $\gamma \in [0, \gamma_c]$, $\omega_k$ become purely imaginary and, therefore, the modes have no oscillatory part and simply blow up or decay away exponentially. These modes correspond to the overdamped modes of a single oscillator, which is of little interest, particularly for sensor applications that require sharp resonances.

**Wireless measurement set-ups.** Our experimental set-up comprised a MEMSbased wireless pressure sensor, inductively coupled to a conventional passive reader or an active reader (a picture of the experimental set-up is shown in Supplementary Note 2). The MEMS varactor is constituted by two circular parallel metal sheets with a diameter of 4mm and an air gap of 100 μm. To simulate variations of internal pressure inside the human eye, the sensor, placed on an XYZ linear translation stage, was encapsulated with epoxy polyamides and connected with an air compressor. A microprocessor-controlled regulator (SMC E/P Regulator) was used to control the internal pressure inside the air cavity of the MEMS varactor. The active reader composed of an –RLC tank was fixed and connected to a VNA (Agilent E5061B). This allows for precise control of the coupling strength $\kappa$ between the MEMS-based pressure sensor and the reader coil. The internal pressure inside the micromachined air cavity of the sensor, as the main physiological parameter of interest, was characterized by tracking the resonance frequency from the measured reflection coefficients. In our experiments, the pressure was varied from 0mmHg to 200mmHg, and the VNA and the pressure regulator were synchronously controlled by the LabVIEW program.

**Data availability statement.** The data that support the plots within this paper and other findings of this study are available from the corresponding author upon reasonable request.

## References

[1] Nopper, R., Niekrawietz, R., and Reindl, L., "Wireless readout of passive LC sensors." *IEEE Trans. Instrum. Meas.* **59**, 2450-2457 (2010).

[2] Collins, C. C., "Miniature passive pressure transensor for implanting in the eye." *IEEE Trans. Biomed. Eng.* **2**, 74-83 (1967).

**Acknowledgements** This work has been supported by the NSF ECCS grant No. 1711409 (to P. Y. C.), Air Force Office of Scientific Research, the Welch Foundation with grant No. F-1802 (to A.A.), Army Research Office (ARO) Grant No. W911NF-17-1-0481 (to R.E.). Device fabrication was carried out in the Nano Fabrication Service Core (nFab) at the Wayne State University.

**Author Contributions** M.S., M.H., and Q.C. designed PT and PTX circuits and performed experimental measurements. M.S., M.H., Q.C., and M.C. designed and fabricated the MEMS pressure sensor. P.Y.C., M.S., and M.C. conceived the experimental concepts. P.Y.C., M.C., R.E.G. and A.A. developed the concepts. P.Y.C. and A.A. planned and directed the research. P.Y.C., R.E.G., and A.A. wrote the manuscript.

**Author Information** Correspondence and requests for materials should be addressed to P.Y.C. (pychen@wayne.edu) and A.A. (aalu@gc.cuny.edu).

**Figures:**

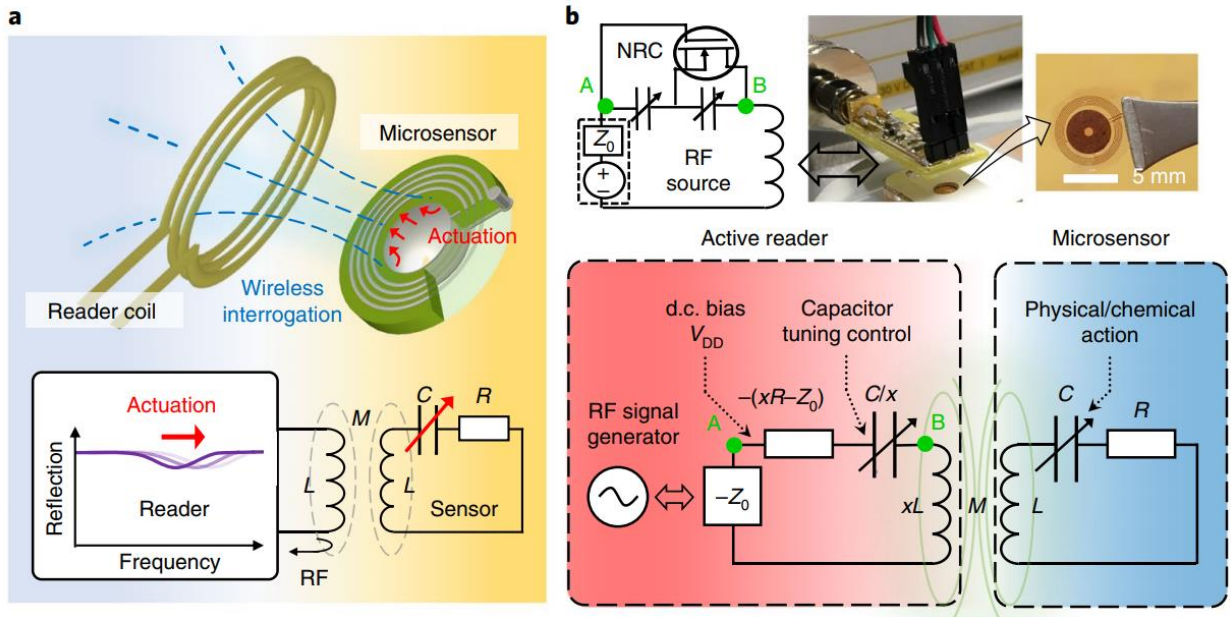

**Figure 1 | Non-Hermitian telemetric sensor system. (a)** Schematics of a typical wireless implantable or wearable sensor system, where a loop antenna is used to interrogate the sensor via inductive (magnetic) coupling. The parameters to be sensed can be accessed by monitoring the reflection coefficient of the sensor, typically based on an *RLC* resonant circuit consisting of a micromachined varactor and inductor. **(b)** Equivalent circuit model for the proposed *PTX*-symmetric telemetric sensor system, where $x$ is the scaling coefficient of the reciprocal-scaling operation $\mathcal{X}$. If $x = 1$, the *PTX* system converges to the *PT*-symmetric case. In the close-loop normal mode analysis, an RF signal generator with a source impedance $Z_0$, connected to the reader, is represented by $-Z_0$. The inset shows the a.c. model for the Colpitts circuit with a positive feedback, which achieves an equivalent negative resistance and an equivalent series capacitance.

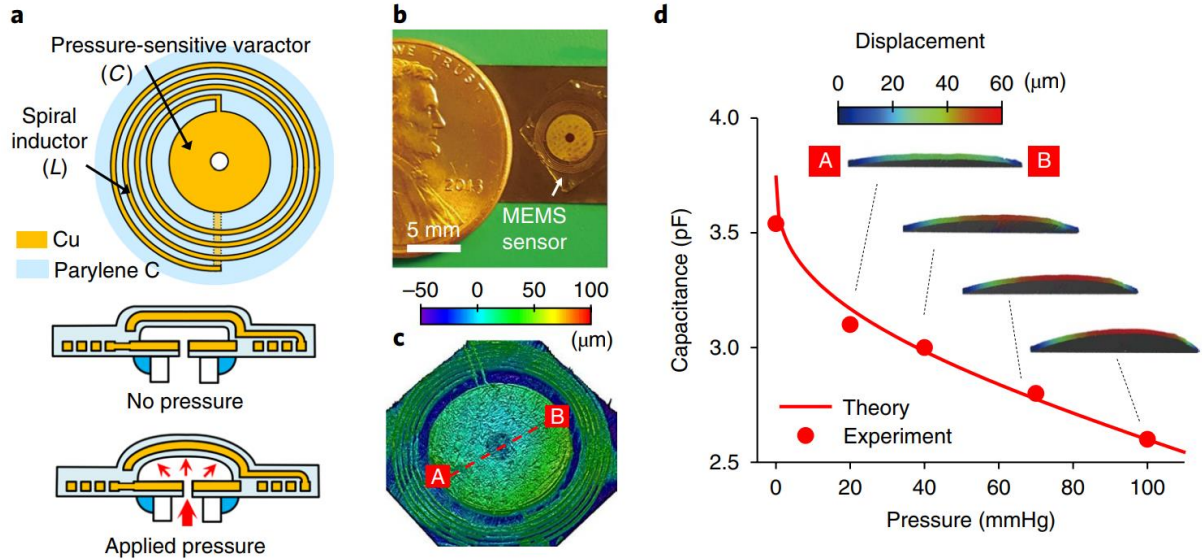

**Figure 2 | MEMS-based wireless pressure sensor.** (**a**) Schematic diagrams of a MEMS-based pressure sensor, which consists of a variable parallel-plate capacitor (*C*) connected in series with a microcoil inductor (*L*), effectively forming a resonant *LC* tank circuit. Increasing the internal pressure by using an air compressor regulator increases the displacement of the upper membrane electrode, thereby reducing the capacitance of the MEMS varactor (**b**) Top view of the microfabricated wireless pressure sensor on a flexible polymer substrate. (**c**) Three-dimensional surface profile of the sensor in **b**, which was measured by scanning white light interferometry (SWLI). (**d**) Measurement (dots) and theoretical (solid line) results for the total capacitance in response to pressure (Supplementary Note 1); the insets show the displacement of the upper membrane electrode measured by SWLI. Due to the cylindrical symmetry of the capacitor, only displacements in the radial direction (from point A to point B in **c**) are shown.

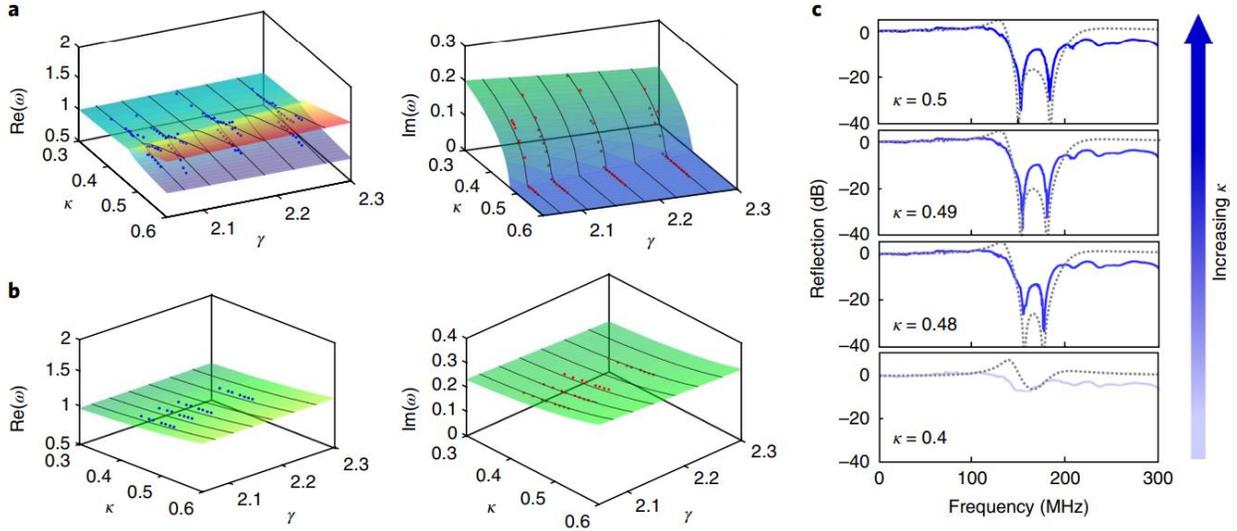

**Figure 3 | Evolution of eigenfrequencies and reflection spectra as a fucntion of the non-Hermiticity parameter $\gamma$ and coupling strength $\kappa$.** **(a,b)** Real (**a**, left) and imaginary (**a**, right) eigenfrequency isosurface normalized by $\omega_0$ in the $(\gamma,\kappa)$ parameter space for a *PT*-symmetric wireless pressure sensor and a conventional passive wireless pressure sensor (**b**), where an active reader and a passive loop antenna are respectively used to interrogate the micromachined sensor in Fig. 2 **(c)** Reflection spectra against the frequency for the *PT*-symmetric wireless pressure sensor with different coupling strengths, showing a transition from the broken *PT*-symmetric phase ($\kappa = 0.4$) to the exact PT-symmetric phase ($\kappa = 0.48$, $0.49$ and $0.5$) when $\kappa$ increases; here $\gamma = 2.26$, corresponding to an applied pressure of 100 mmHg, and $\omega_0 / 2\pi = 180$ MHz. The frequencies and linewidths of the reflection dips in **c** are consistent with the eigenfrequency evolution in **a**. The solid and dashed lines denote experimental data and theoretical results obtained from the equivalent circuit model in Fig. 1.

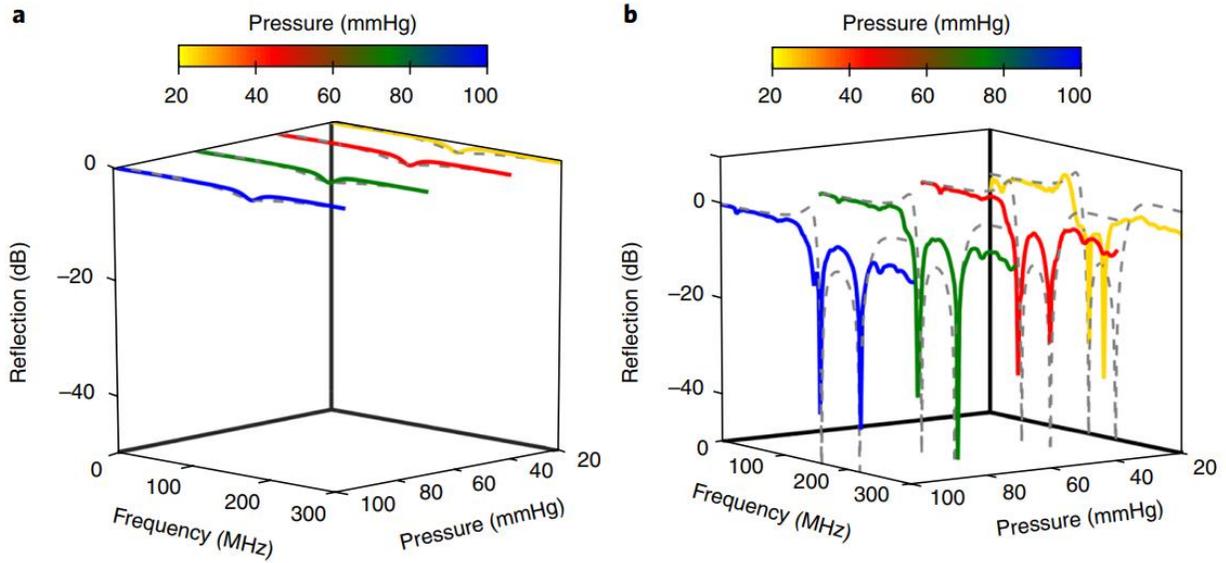

**Figure 4 | Pressure-induced spectral changes for conventional and *PT*-symmetirc telemetric sensors. (a,b)** The magnitude of the reflection coefficient for the MEMS-based pressure sensor (Fig. 2) interrogated by the conventional passive loop antenna **a** and the active reader realizing a *PT*-symmetric dimer **b**, under different applied pressures. The solid and dashed lines denote experimental data and theoretical results obtained from the equivalent circuit models.

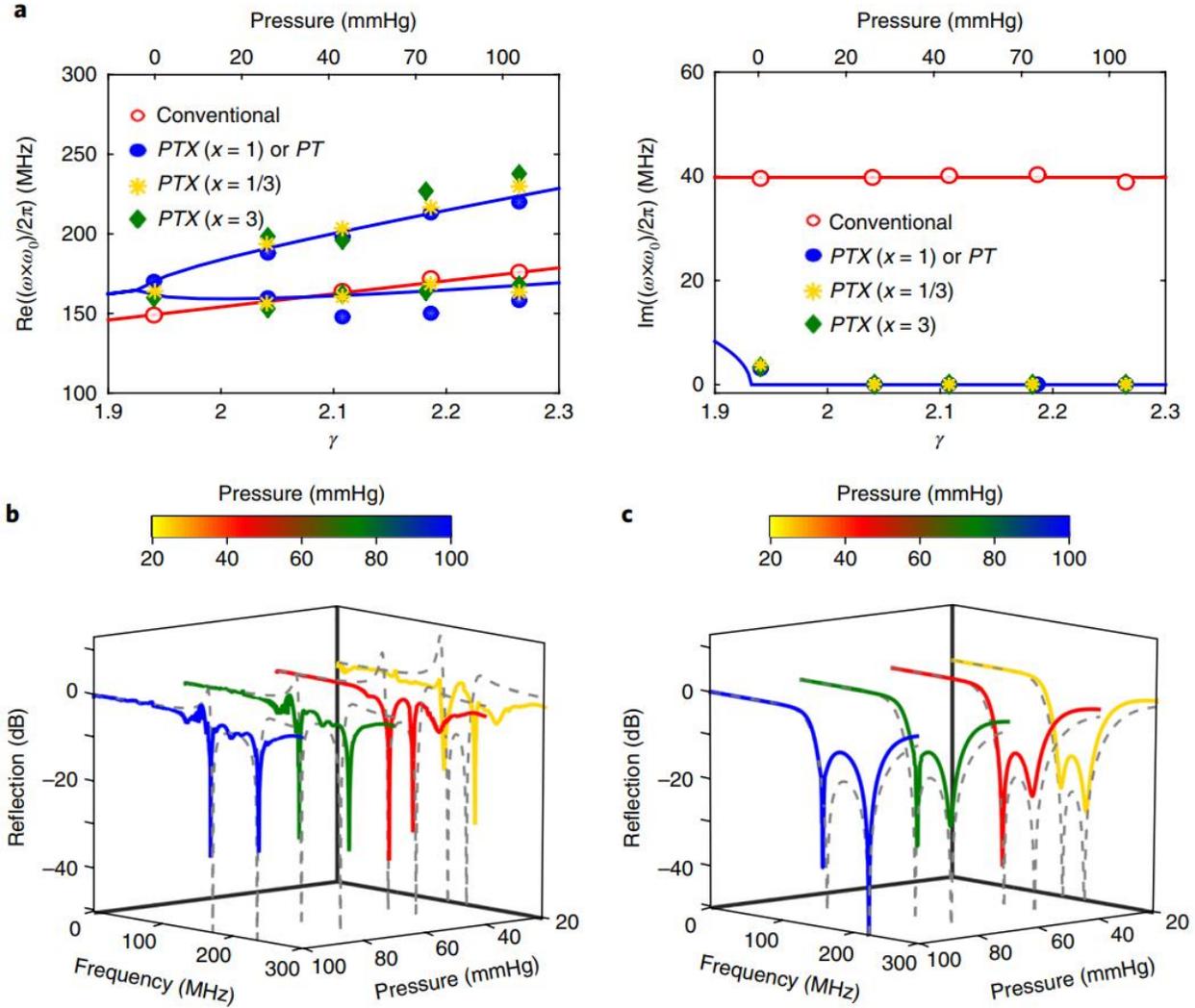

**Figure 5 | Evolution of the eigenfrequencies and reflection spectra for PTX-symmetric telemetric sensors. (a)** Real (left) and imaginary (right) eigenfrequency as a function of the non-Hermiticity parameter $\gamma$ for the fully passive (red open circles; Fig. 4a), *PT*-symmetric (blue dots; Fig. 4b) and *PTX*-symmetric (green and yellow symbols; **b** and **c**) telemetric pressure sensors. The solid lines denote theoretical predictions (see Methods). **(b,c)** The magnitude of the reflection coefficient for the *PTX*-symmetric telemetric sensor systems; here, the scaling coefficients $x$ used in **b** and **c** are 3 and 1/3. The solid and dashed lines in **b** and **c** denote the experimental data and theoretical predictions.

# Supplementary information

**S1. Design and Characterization of MEMS-Based Pressure Sensor**

*S1.1 Design of MEMS-Actuated Capacitive Pressure Sensor*

A typical passive pressure sensor contains an *LC* resonator, including a pressure-tuned parallel-plate capacitor and a planar micro-coil inductor. Such device architecture has been widely adopted for pressure sensors in many medical, industrial, automotive, defense and consumer applications [1]-[9]. Assuming no fringe effect, the capacitance is given by:

$$C = \varepsilon_0 \varepsilon_r \frac{A}{d}, \tag{S1}$$

where $\varepsilon_r$ is the relative permittivity, $\varepsilon_0$ is free space permittivity, $A$ and $d$ are the area of two capacitor electrodes and the separation distance between them (when no pressure is applied). As schematically shown in Figure 2 in the main text, the MEMS varactor includes a movable upper electrode and a stationary lower electrode, which are separated by a variable air gap ($\varepsilon_r = 1$). The lower electrode is fixed to the substrate and has a small drain hole connected to the compressor through a sealed tube. Therefore, the pressure inside the encapsulated air cavity can be controlled by a pressure regulator. As the internal pressure increases, the upper electrode is gradually bent upward such that the total capacitance of the MEMS varactor is varied. The maximum displacement of the movable upper electrode $\Delta d$, as a function of pressure $P$ and electrode's material parameters (Young's modulus $E$ and Poison ratio $v$), can be calculated using the Euler–Bernoulli theory [5], leading to:

$$\Delta d = \frac{3Pa_0^4(1-v^2)}{16Et^3} \frac{1}{1+0.448\left(\frac{d}{t}\right)^2}, \tag{S2}$$

where $a_0$ and $t$ represent the radius and thickness of the circular metallic plates. Consider the pressure-driven displacement, the capacitance can be calculated by conducting the surface integral over the metallic disk:

$$C' = 2\pi\varepsilon_r\varepsilon_0 \int_0^{a_0} \frac{r}{d+\delta d(r)} dr, \tag{S3}$$

where $d(r)$ is the function of deflection depending on the radial position of the membrane. Under an internal pressure, $C'$ can be approximately expressed as a function of $\Delta d$ [5]:

$$C' = C \frac{\sqrt{\frac{\Delta d}{d}}}{\tanh^{-1}\sqrt{\frac{\Delta d}{d}}}. \tag{S4}$$

For most commonly used copper electrodes, important material parameters are: $E = 117$ GPa and $v = 0.33$ [10]. In our design, the two copper disks have the same radius $a_0 = 2$ mm and are initially separated by an air gap $d = 100$ μm.

Supplementary Figure 1 shows the theoretical and measurement results for the maximum displacement of the movable upper electrode as a function of the applied pressure. The scanning white-light interferometry (SWLI) was used to determine the maximum displacement. It is seen from Supplementary Figure 1 that the experimental results agree with the theory quite well, confirming the validity of Eq. (S2). As can be expected, the displacement of upper electrode increases with increasing the applied pressure, which, in turn, reduces the total capacitance.

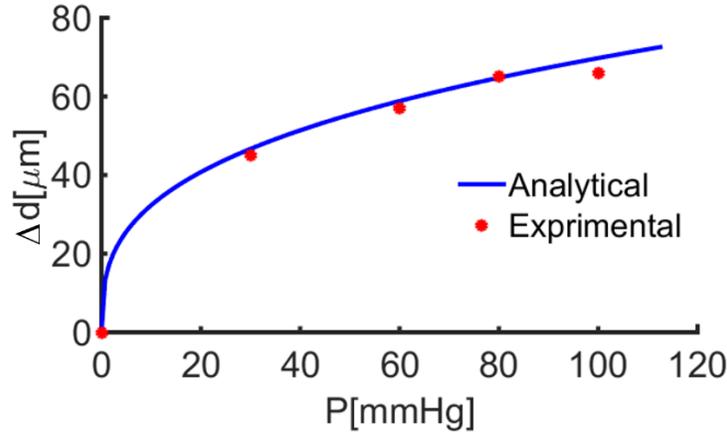

*Supplementary Figure 1. Maximum displacement of the movable electrode against applied pressure for the MEMS varactor in Figure 2 of the main text.*

To characterize the practical capacitance and the effective resistance of the sensor, we first used an external coil to contactlessly read the sensor, and then analyzed the reflection responses to retrieve lumped-element parameters in the equivalent circuit. In our characterizations, we first characterized an individual micro-coil (without loading the capacitor) for knowing its inductance value, as well as the mutual inductance between two tightly coupled micro-coils. Once the impedance of the micro-coil is known, the capacitance of the complete sensor as a function of applied pressure can be retrieved by fitting experiment data with the equivalent circuit model. From the complex reflection coefficient, the effective resistance of the sensor can also be retrieved, which is found to be almost invariant under different pressures (~150 Ω). Supplementary Figure 2 presents theoretical and experimental values of capacitance of the MEMS varactor; here capacitance as a function of pressure was calculated using Eqs. (S2)-(S4). The experimental and theoretical results exhibit good agreement, despite slight differences due to fringing effects and microfabrication imperfections. It is seen from Supplementary Figure 2 that the sensor's capacitance decreases with increasing the applied pressure, due to the enlarged air gap $\Delta d$ (Supplementary Figure 1).

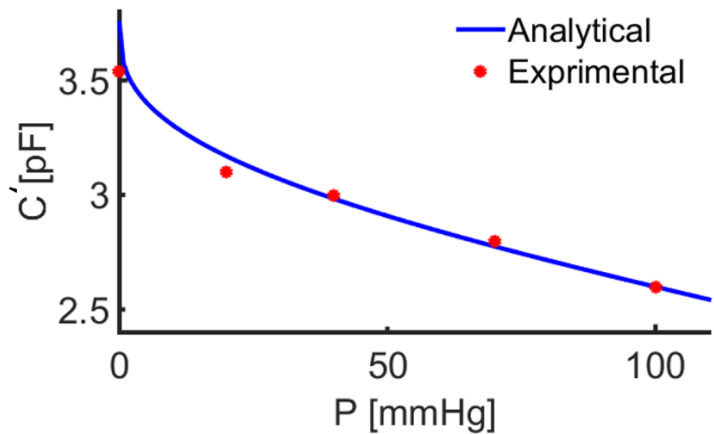

***Supplementary Figure 2.*** *Capacitance against the applied pressure for the MEMS varactor in Figure 2 of the main text.*

*S1.2 Design of Microcoil Inductor*

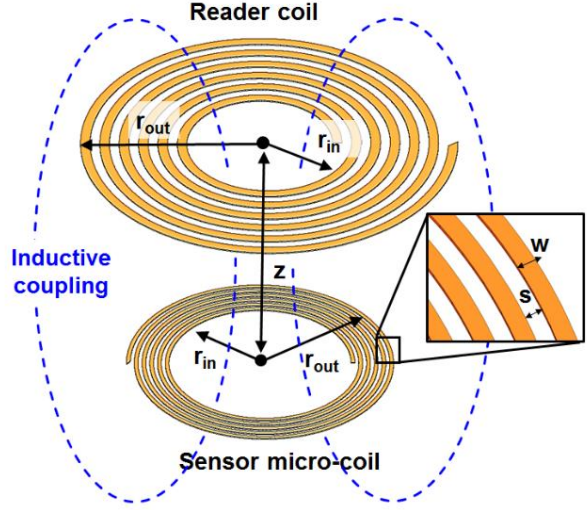

***Supplementary Figure 3.*** *Configurations and physical parameters of planar coils used in the reader and the sensor.*

The self-inductance of the planar micro-coil inductor in Supplementary Figure 3 can be derived from the ratio between the magnetic flux and current, which has an approximate expression as [11]:

$$L = \frac{\mu_0 N^2 d_{avg}}{2}\left[\ln\left(\frac{2.46}{\varphi}\right) + 0.2\varphi^2\right], \tag{S5}$$

where $\mu_0$ is the free space permeability, $N$ is number of turn, $d_{avg} = 2r_{in} + N\times(s+w)$ is the average diameter of spiral coil, $2r_{in}$ is inner diameter of spiral coil, $w$ and $s$ are width and spacing of the coil, and $\varphi = N\times(s+w)/[d_i + N\times(s+w)]$ is the filling ratio. We have applied Eq. (S5) to design the reader/sensor micro-coils. For example, the inductance values and important design parameters for micro-coils used in the *PT*-symmetric sensor (Figure 4b) are summarized in Table 1.

*Table 1. Physical parameters for IOP sensor and reader.*

|        | $L$ [µH] | $N$ | $s$ [mm] | $w$ [mm] | $r_{in}$ [mm] |
|--------|----------|-----|----------|----------|---------------|
| Sensor | 0.3      | 5.5 | 0.075    | 0.075    | 2.4           |
| Reader | 0.28     | 6   | 0.25     | 0.25     | 2             |

The mutual inductance for two filamentary currents $i$ and $j$ can be computed using the double integral Neumann formula [12]:

$$M_{ij} = \frac{\mu_0}{4\pi} \int_{C_i} \int_{C_j} \frac{1}{|R_{ij}|} d\vec{l}_i \cdot d\vec{l}_j, \tag{S6}$$

where $R_{ij}$ represents the distance between metallic lines, which has a relation with the radius of each coil and the central distance between them. The calculation of total mutual inductance for coils with multiple turns is possible with the summation of the separate mutual inductance of each current filament:

$$M = \rho \sum_{i=1}^{N_R} \sum_{j=1}^{N_S} M_{ij}, \tag{S7}$$

where $i$ ($j$) represents the $i$-th ($j$-th) turn of micro-coil on the reader (sensor) side, $\rho$ is the shape factor of planar coil [12], and $M_{ij}$ is the mutual inductance between the loops $i$ and $j$, which are given by:

$$M_{ij} \approx \frac{\mu_0 \pi a_i^2 b_j^2}{2(a_i^2 + b_j^2 + z^2)^{3/2}}, \tag{S8}$$

where $z$ is the central distance between two micro-coils, $a_i = r_{o,R} - (N_i - 1)(w_R + s_R) - w_R/2$, $b_j = r_{o,S} - (N_j - 1)(w_S + s_S) - w_S/2$, $N_i$ ($N_j$) represents the $i$-th ($j$-th) turn of reader (sensor) coil, $r_o$ is the outer radius of the microcoil, and the subscript $R$ ($S$) represents reader (sensor). Finally, the coupling coefficient between the reader and sensor micro-coils is given by $\kappa = M/\sqrt{L_R L_S}$, where $L_R$ is the reader coil inductance and $L_S$ is the sensor coil inductance. In our designs, we first characterized the self-inductance of each individual coil using the analytical formula of Eq. (S5), which has been verified with the full-wave simulation [13]. Then, the total mutual inductance between two micro-coils was calculated using the analytical formula of Eqs. (S7)-(S8) and the result was confirmed by the full-wave simulations. In our designs, the coupling coefficient $\kappa$ is in the range of 0 to 0.5.

## S1.3 Design of Negative Resistance Converter (NRC)

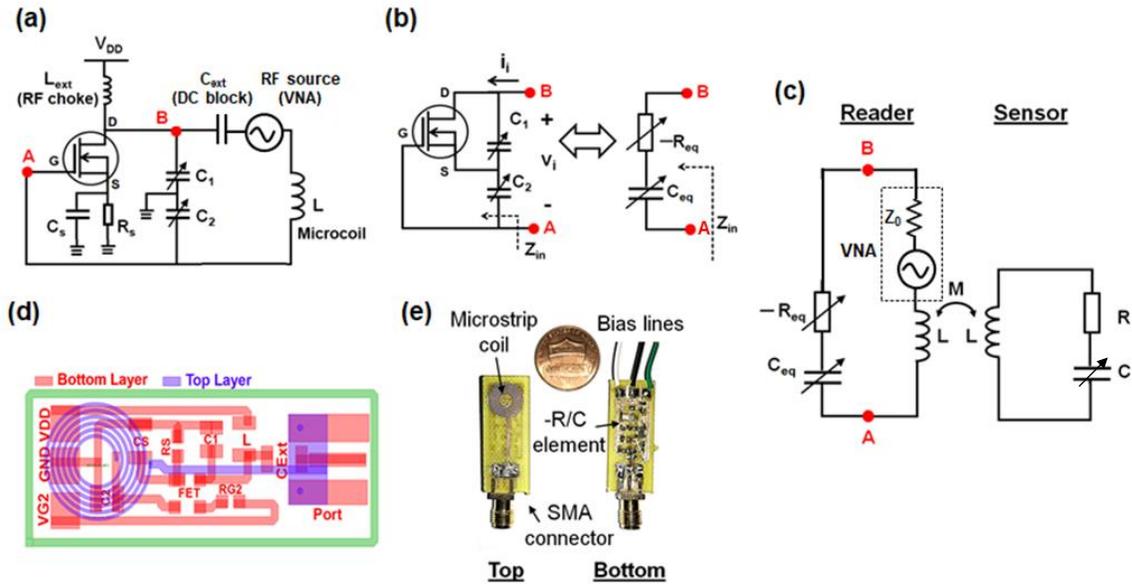

***Supplementary Figure 4.** (a) Schematics of the reader circuit for the PT/PTX-symmetric telemetric sensors, consisting of a negative resistance converter (Colpitts oscillator) connected to a microcoil inductor, fed by a RF source (vector network analyzer; VNA). (b) Input impedance of an open-circuited Colpitts-type circuit $Z_{in}=v_i/i_i$, without connecting to any reactive element. The complex input impedance can be decomposed into a series combination of an equivalent negative resistance $-R_{eq}$ and an equivalent capacitance $C_{eq}$. (c)Equivalent circuit model for the PT/PTX-symmetric telemetric sensor system, in which the reader is a series $-RLC$ tank where $-R$ and C are contributed by the Colpitts-type oscillator. (d) Layout and (e) fabricated PCB-based active reader used in the PT-symmetric sensor.*

To build the *PT-/PTX*-symmetric electronic circuit, it requires a negative resistor ($-R$), realized using a negative resistance converter (NRC) at high frequencies. An active NRC could pull in power to the circuit, rather than dissipating it like a passive resistor. Supplementary Figure 4a shows the circuit diagram of our NRC, inspired by the design of Colpitts-type oscillator [14]- [21]. This NRC as an active lumped resistor may provide stable and almost non-dispersive negative resistance over a broad frequency range. The negative resistance can be a series or a parallel element, depending on how the circuit is designed, i.e., a series (parallel) circuit model is usually used for voltage-controlled negative resistance oscillators (current-controlled negative conductance oscillators) [22]. For example, in Ref. [23], a one-port op-amp inverting circuit operating at KHz frequencies, equivalent to a parallel negative resistance, was used to demonstrate a *PT*-symmetric system based on parallel −*RLC* and *RLC* tanks. In the circuit analysis, it is common to model the Colpitts- or Hartley-type oscillator with positive feedback as a negative resistor (−*R*). One method of oscillator analysis is to determine its input impedance, neglecting any external reactive component at the input port, as shown in Supplementary Figure 4b. For the Colpitts circuit configuration in Supplementary Figure 4b, the complex input impedance ($Z_{in} = v_i/i_i$) looking into the points A and B can be derived as [14]-[21]:

$$Z_{in(AB)} = -\frac{g_m}{\omega^2 C_1 C_2} + i\left(\frac{1}{\omega C_1} + \frac{1}{\omega C_2}\right), \tag{S9}$$

where $g_m$ is the transconductance of the field-effect transistor (FET); here we assume that the $g_m \gg \omega C_{gd}, \omega C_{gs}$ ($C_{gd}$ and $C_{gs}$ are the gate-drain capacitance and gate-source capacitances), which is approximately valid at moderately low frequencies (e.g., VHF band in this paper). As a result, the input impedance looking into the points A and B is equivalent to a series −*RC* circuit consisting of a negative resistance $-R_{eq}$ and an equivalent capacitance $C_{eq}$, as shown in Supplementary Figure 4b [1]-[6]:

$$-R_{eq} = -\frac{g_m(V_{bias})}{\omega^2 C_1 C_2} \quad \text{and} \quad C_{eq} = \frac{C_1 C_2}{C_1 + C_2}. \tag{S10}$$

By connecting the input port to an inductor, a positive feedback oscillator can be made by controlling the open-loop and feedback gains at the resonance frequency. As known form Eq. (S10), the negative resistance can be increased by using larger values of transconductance and smaller values of capacitance. If the two capacitors are replaced by inductors, the circuit becomes a Hartley oscillator, whose input impedance becomes the −*RL* combination.

According to Eqs. (S9) and (S10), the effective resistance is controlled by the transistor's transconductance, readily adjusted by DC offset voltages. The RF transistors used here have high cutoff frequencies up to several GHz, ensuring the minimum parasitic effects and the stability of circuit. The effective capacitance is determined by the two lumped capacitances $C_1$ and $C_2$, which could be contributed by the voltage-controlled varactors such that the effective capacitance of the $-RLC$ tank is tunable. If a microcoil inductor is connected to the input of the Colpitts oscillator (points A and B in Supplementary Figures 4b and 4c), a series $-RLC$ tank can be realized if that the AC source is connected in series to the inductor, as can be seen in Supplementary Figure 4c. Supplementary Figures 4d and 4e show the circuit layout and the fabricated printed circuit board (PCB) for the active reader used in *PT*-symmetric system (Figure 4b), respectively. This active reader consists of the voltage-tuned NRC (Supplementary Figure 4a), which are connected in series to a planar coil, forming the $-RLC$ tank.

The effective impedance of the NRC can be retrieved from the measured reflection coefficient of an isolated series $-RLC$ tank connected to the vector network analyzer (VNA), by decomposing the contribution of the coil inductance. An individual $-RLC$ tank can allow the reflected RF signal to have larger amplitude than the incident one, namely the steady-state reflection gain is achieved. However, in experiments the reflection cannot be infinitely large because all transistors and electronic components have maximum operating voltage/current ranges, large-signal effects, and inherent nonlinearities. In a similar sense, although in theory a pole could arise in a $-RLC$ tank, an ever-growing eigenmode (charge/charge flow) is never achieved due to the above-mentioned nonlinear effects in real-world electronic devices. A more detailed equivalent circuit of the Colpitts-type NRC is shown in Supplementary Figure 5a, which includes also a shunt inductance $L_p$ and a parasitic capacitance $C_p$ [22]. We note that, at sufficiently low frequencies, $C_p$ has a high RF impedance $Z_c = i/\omega C_p$ (acting like a low-pass filter or an open circuit), while $L_p$ has a low RF impedance $Z_l = -i\omega L_p$ (acting like a short circuit). Therefore, the parasitic effect may be minimized if the operating frequency is moderately low, well below the transistor's cutoff frequency ($f_T$) and maximum frequency ($f_{max}$). Supplementary Figure 5a presents the experimental (solid) and simulated (dashed) reflection spectra of the Colpitts-NRC shown in Supplementary Figure 4, under different DC bias conditions. In our simulations, the equivalent resistive and reactive values in the circuit model, as shown in Supplementary Figures 5b and 5c, were extracted from the measured reflection coefficients by using the numerical optimization. From Supplementary Figure 5a, a good agreement is found between the experimental and simulation results. Here, we also present the reflection spectra for the equivalent circuit in Supplementary Figure 5a without the parasitic capacitance (dotted). The results show no significant difference in the frequency range of interest, when compared to those obtained from experiments and the full equivalent circuit model.

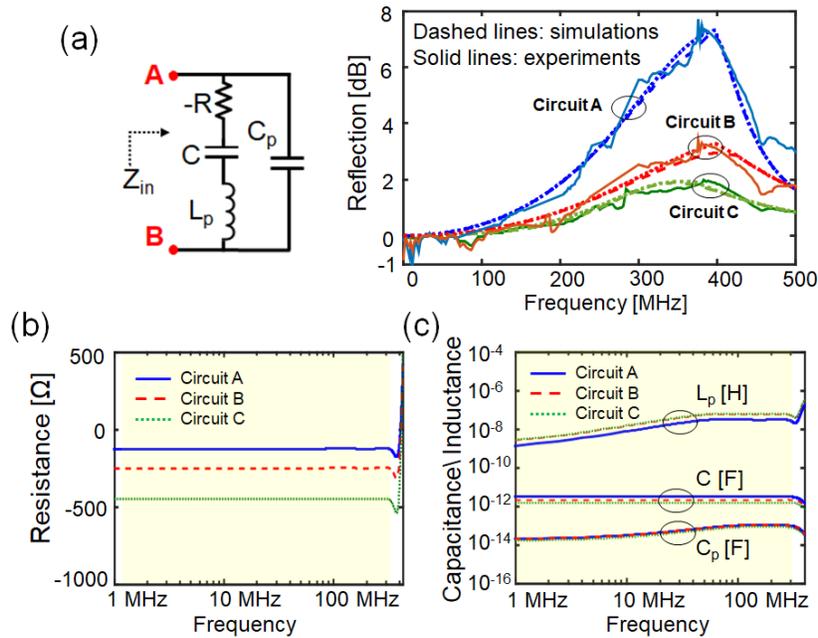

***Supplementary Figure 5.*** *(a) Reflection spectra of the NRC versus frequency under different biasing conditions; here, solid and dashed lines represent the experimental and simulation results, and dotted lines represent the simulation results without considering the parasitic capacitance. (b) Equivalent resistance and (c) equivalent capacitance and parasitic components for the NRC in (a). The highlighted areas show the frequency range of interest, where the values of negative resistance and capacitance are nearly constant. The effects of $L_p$ and $C_p$ are negligible if the operating frequency is much lower than the cutoff frequency of the transistor.*

As a result, for our initial analysis, parasitic elements and device nonlinearities are ignored. In the frequency range of interest, the experimentally measured input impedance can be decomposed into a series combination of a negative resistance and a capacitance. This simplified model shows an acceptable comparison with experimental results, as can be seen in Supplementary Figure 5a. It is clearly seen from Supplementary Figure 5b that that negative resistance can be tuned by adjusting the DC offset voltage and their values are nearly invariant at low frequencies.

**S2. Microfabrication and Characterization of the Wireless Pressure Sensor**

*S2.1 Fabrication of Wireless Pressure Sensors by the MEMS processes*

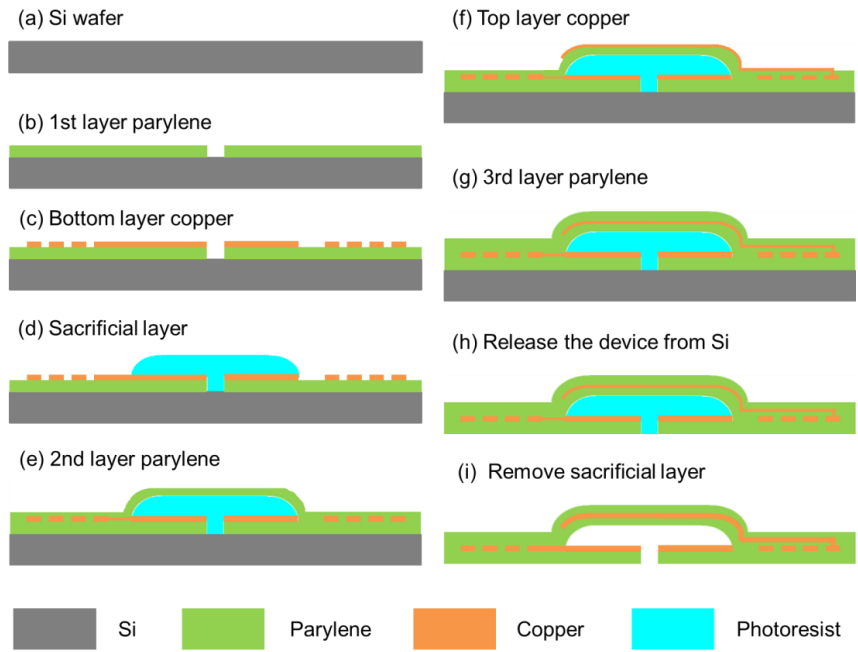

***Supplementary Figure 6.*** *Schematics of fabrication processes for the MEMS-based intraocular pressure sensor in Figure 2 in the main text.*

Supplementary Figure 6 presents the fabrication flow of the MEMS-based intraocular pressure (IOP) sensor in Figure 2. In step (a), the silicon (Si) wafer was first cleaned following the standard RCA cleaning process. In step (b), an 8 μm-thick parylene layer was deposited using the thermally activated chemical vapor deposition (CVD) method in the Specialty Coating Systems (SCS PDS 2010). In this parylene layer, the sensor's backside was etched to form a pressure access hole with diameter of 0.8 mm by using the oxygen reactive ion etching (RIE, DryTech RIE 184).

In step (c), a 3 μm-thick copper (Cu) film was deposited using the electron beam evaporation (Temescal Model BJD-1800 e-beam evaporator). Standard photolithography was used to pattern the copper and form the coil inductor and the capacitor pad. In steps (d)-(e), a sacrificial photoresist layer was patterned by the lithographic method, following the coating of the second parylene layer that has a thickness of 4 μm. In step (f), the second Cu layer was deposited and patterned by the lithographic method; here the top and bottom metallic structures were connected by a Cu interconnect patterned by oxygen RIE. In step (g), the third parylene with a thickness of 8 μm was coated as protection layer. After that, the device was released from Si substrate using the KOH solution. In the last step, (h), the sacrificial layer was removed in acetone solution with the critical point dryer (CPD) (Tousimis 931). Finally, the microfabricated IOP sensor was made with a flexible air cavity that can be actuated by the internal pressure, as shown in (i).

## S.2.2 Measurement Setup for the MEMS Wireless Pressure Sensors

Supplementary Figure 7 shows our wireless measurement setup, which comprises a MEMS-based pressure sensor, inductively coupled to a passive or active reader (interrogator). The MEMS-based pressure sensor consists of a variable capacitor (varactor) functioning as a transducer, connected in series to a planar microcoil inductor. In the equivalent circuit diagram, the pressure sensor itself stands for a *RLC* tank, where the applied pressure mechanically deforms the MEMS varactor and therefore varies the sensor's natural frequency. The pressure sensor was encapsulated with epoxy polyamides and connected with an air compressor. A microprocessor-controlled regulator (SMC E/P Regulator) was used to control the internal pressure inside the air cavity of MEMS varactor. The sensor was fixed on a XYZ linear translation stage and the active reader composed of –*RLC* tank was connected to vector network analyzer (VNA: Agilent E5061B). This setup allows for precise control of the coupling strength $\kappa$ between the MEMS sensor and the reader coil.

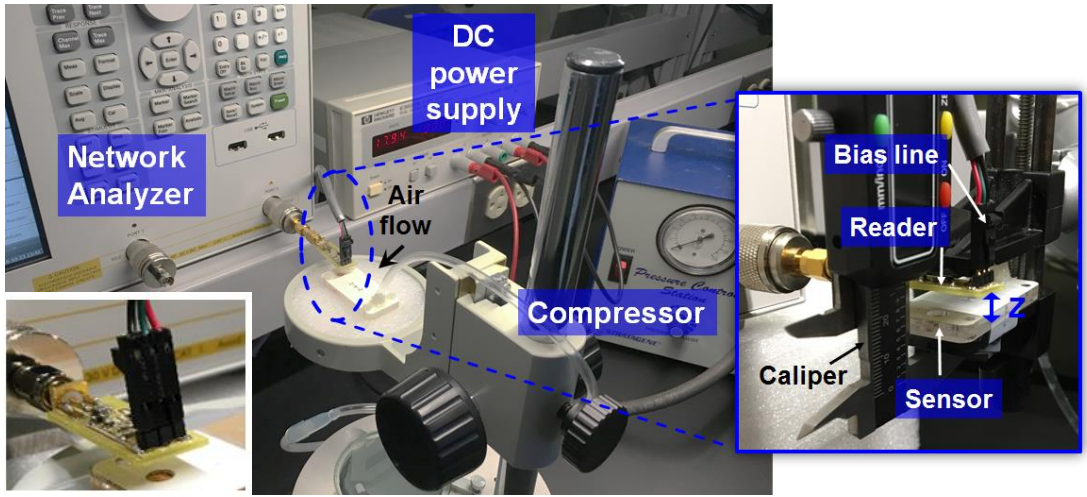

***Supplementary Figure 7.*** *Measurement Setup for the MEMS wireless pressure sensor.*

Supplementary Figure 8 shows the theoretical and experimental results for eigenfrequency variations of the *PT*-symmetric wireless pressure sensor system, where the capacitance of the microfabricated pressure sensor is changed with respect to the pressure-induced displacement; here an equivalent resistance of $150\,\Omega$ was measured for the sensor, and the inductance of the sensor's micro-coil is about 300 nH. When the sensor's capacitance is varied (Supplementary Figure 2), the effective capacitance of the active reader should also be tuned accordingly to maintain the *PT*-symmetry. This can be achieved by precisely controlling the DC offset voltage of varactors in the reader circuit. To make a fair comparison, we also studied the conventional wireless pressure sensor system, where a passive external coil (the same as the one used in the active reader) was used to read the microsensor. As can be seen in Supplementary Figure 8, the *PT*-symmetric telemetric sensor system can provide a larger resonance frequency shift in response to pressure-driven capacitance variations. We note that a *PTX*-symmetric sensor would display the same sensitivity because the *PT* and *PTX* systems share the same eigenspectrum, as discussed in the main text.

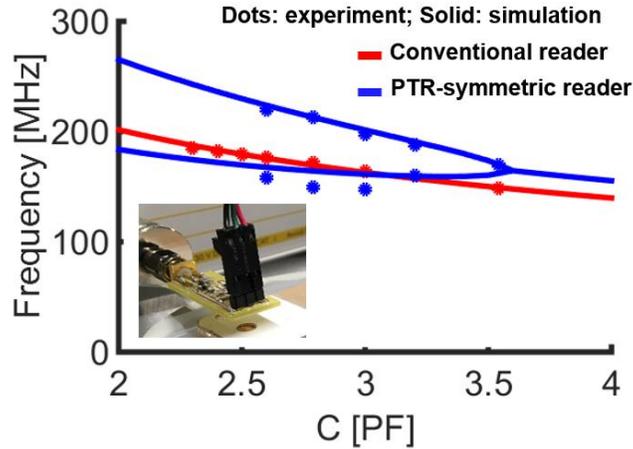

***Supplementary Figure 8.*** *Variations of eigenfrequencies with the pressure-driven capacitance for conventional and PT-symmetric wireless pressure sensor systems. The pressure corresponding to the specific capacitance can be found in Supplementary Figure 2. We note that a PTX-symmetric sensor displays the same frequency response because PT and PTX systems share the same eigenspectrum.*

### S3. Analysis of PTX-Symmetric Electronic Systems

*S3.1 PTX-Symmetric Circuits in the Parallel Configuration*

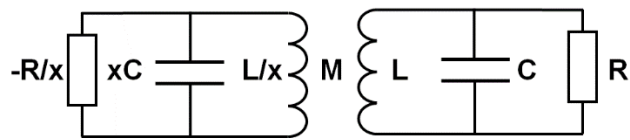

***Supplementary Figure 9 .*** *PTX-symmetric electronic system realized with the parallel-circuit configuration.*

The passive *LC* wireless sensors are also commonly designed and modeled using an equivalent, *parallel RLC circuit (excited by an impressed current source). We note that the concept of PTX-symmetry can in principle be applied to different types of series and parallel circuits, and possibly their complex combinations.* Supplementary Figure 9 shows a *PTX*-symmetric circuit formed by the parallel $-RLC$ and $RLC$ tanks, which communicate through the inductive coupling. Such a system is invariant under the parity transformation $\mathcal{P}$ ($q_1 \leftrightarrow q_2$), time-reversal transformation $\mathcal{T}$ ($t \to -t$), and reciprocal scaling $\mathcal{X}$ ($q_1 \to x^{1/2} q_1$, $q_2 \to x^{-1/2} q_2$), where $q_1$ ($q_2$) corresponds to the charge stored in the capacitor in the parallel $-RLC$ ($RLC$) tank. Its $PT$-symmetric counterpart with $x = 1$ have been experimentally demonstrated in Ref. [23], in which a shunt negative resistor was realized using the op-amp inverting circuit. According to the Kirchoff's law, the Liouvillian $\mathcal{L}$ and the effective Hamiltonian $H$ of the $PTX$-symmetric circuit in Supplementary Figure 9 can be derived as:

$$H = i\mathcal{L} \text{ and } \mathcal{L} = \begin{pmatrix} 0 & 0 & 1 & 0 \\ 0 & 0 & 0 & 1 \\ -\dfrac{1}{1-\kappa^2} & \dfrac{1}{\sqrt{x}}\dfrac{\kappa}{1-\kappa^2} & \gamma & 0 \\ \sqrt{x}\dfrac{\kappa}{1-\kappa^2} & -\dfrac{1}{1-\kappa^2} & 0 & -\gamma \end{pmatrix} \tag{S11}$$

where $\omega_0 = 1/\sqrt{LC}$, the coupling strength between the active and passive tanks $\kappa = \sqrt{x}M/L$, the non-Hermiticity parameter $\gamma = R^{-1}\sqrt{L/C} = (|-R|/x)^{-1}\sqrt{(L/x)/(xC)}$, and all frequencies are measured in units of $\omega_0$. The effective Hamiltonian is non-Hermitian (i.e., $H^\dagger \neq H$) and commutes with $\mathcal{PTX}$; here $\mathcal{P}$, $\mathcal{T}$, and $\mathcal{X}$ are defined in Eq. (2) in the main text. The Hamiltonian and eigenmodes of this $PTX$ system can be linked to those of its $PT$ counterpart $(H', \Psi')$ through the similarity transformation $H = S^{-1} H' S$ and $\Psi = S^{-1} \Psi'$, where $S$ is an invertible 4-by-4 matrix $S = \mathbf{1} \otimes \zeta$ and $\zeta = \begin{pmatrix} x^{1/2} & 0 \\ 0 & 1 \end{pmatrix}$. As a result, the $PTX$ and $PT$ systems share the same eigenfrequencies, given by:

$$\omega_{1,2,3,4} = \pm\sqrt{\dfrac{2-\gamma^2(1-\kappa^2) \pm \sqrt{4\kappa^2 - 4\gamma^2(1-\kappa^2) + \gamma^4(1-\kappa^2)^2}}{2(1-\kappa^2)}}, \tag{S12}$$

which is found to be independent of *x*. Such results are consistent with our previous findings on the series –*RLC*/*RLC* dimer satisfying the *PTX*-symmetry. The scaling coefficient *x* plays a role in controlling the linewidth of the resonance. Therefore, the *PTX*-symmetry concept can also be exploited to improve the *Q*-factor and sensitivity of a wireless resonant sensor based on a parallel *RLC* circuit model.

## S3.2 Reflectionless Property and Impedance Matching

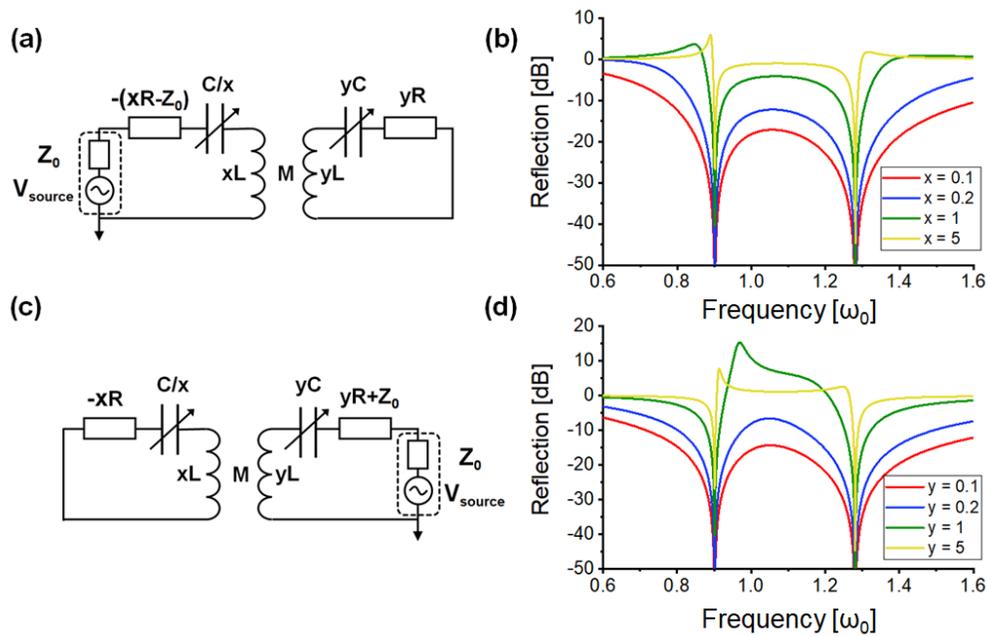

***Supplementary Figure 10.*** *(a) PTX-symmetric circuits with the RF excitation source connected to the active −RLC tank. (b) Reflection spectrum for the single-port circuit in (a), under different values of x. (c) PTX-symmetric circuits with the RF excitation source connected to the passive RLC tank; the circuits in (a) and (c) share the same eigenfrequencies, as they represent the same type of −RLC/RLC dimer in the coupled-mode analysis. (d) Reflection spectrum the single-port circuit in (a), under different values of y. In the PTX-symmetric circuits, the resonant frequencies remain constant, while the bandwidth (or Q-factor) can be tailored by varying the scaling coefficient x or y.*

Supplementary Figure 10a considers a generalized *PTX*-symmetric circuit that is invariant under the $\mathcal{PTX}$ transformation. Here, $\mathcal{X} = \mathbf{1} \otimes x_0$ and $x_0 = \begin{pmatrix} (x/y)^{1/2} & 0 \\ 0 & (x/y)^{-1/2} \end{pmatrix}$, which yield $q_1 \to (x/y)^{1/2} q_1$ and $q_2 \to (x/y)^{-1/2} q_2$. In this case, both active and passive tanks have the same non-Hermiticity parameter as $\gamma = (x|-R|)^{-1}\sqrt{(xL)/(C/x)} = (yR)^{-1}\sqrt{(yL)/(C/y)}$. For this coupled circuit, the input impedance looking into the $-RLC$ tank from the RF generator end can be derived as:

$$Z_{in} = Z_0 \frac{\omega^2 - i\omega\gamma(\omega^2-1) - x\left[\omega^2 - \gamma^2\left(2\omega^2 + \omega^4(\mu^2-1)-1\right)\right]/\eta}{\omega^2 - i\omega\gamma(\omega^2-1)}, \tag{S13}$$

where $\omega$ is the angular frequency, the generator impedance $Z_0 = \eta R \ [\Omega]$, and $\eta$ is the impedance normalization factor. In the single-port measurement, the information is encoded in the reflection coefficient at the input port, which can be written as:

$$\Gamma = (Z_{in} - Z_0)/(Z_{in} + Z_0). \tag{S14}$$

It is interesting to note that the input impedance and the reflection coefficient are independent of *y* used in the *RLC* tank. This could enable more flexibility in the sensor design when compared with the traditional *PT*-symmetric setup. The input impedance and reflection coefficient of the *PT*-symmetric telemetry system are obtained by setting $x = 1$ in Eq. (S13),(S14). In the exact *PT*-/*PTX*-symmetric phase, the eigenfrequencies are real, corresponding to the dips in the reflection spectrum. From the RF circuit viewpoint, the reflectionless property is due to the perfect impedance matching, namely $Z_{in} = Z_0$ at the eigenfrequencies (resonance frequencies), leading to $\Gamma = 0$.

Supplementary Figure 10b shows the reflection spectrum for the *PTX*-symmetric circuits in Supplementary Figure 10a, under different values of *x*; here $\gamma = 2.5, \mu = 2.5, \eta = 0.2,$ and *y* is an arbitrary positive real number (because $Z_{in}$ and $\Gamma$ are independent of *y*). The *PT*-symmetric system is obtained when $x = y = 1$. It can be seen from Supplementary Figure 10b that for different *PTX*-symmetric systems, the reflection coefficient is always zero at the given eigenfrequencies, while the resonance linewidth can be tuned by varying the scaling coefficient *x*. Most importantly, the *Q*-factor, which is inversely proportional to the resonance bandwidth, increases with increasing the value of *x*, as has been demonstrated in our telemetry experiments (Figure 5). As opposed to an active reader, a passive reader with $x \leq \eta$ can exhibit low reflection over a broad bandwidth, which could be of interest for applications that require large amounts of bandwidth, such as the high-speed communication.

Supplementary Figure 10c considers the second case, in which the RF input port is connected to the passive *RLC* tank (e.g., an active sensor interrogated by a passive reader). In this scenario, the input impedance can be derived as:

$$Z_{in} = Z_0 \frac{\omega^2 + i\omega\gamma(\omega^2 - 1) + y\left[\omega^2 - \gamma^2\left(2\omega^2 + \omega^4(\mu^2 - 1) - 1\right)\right]/\eta}{\omega^2 + i\omega\gamma(\omega^2 - 1)}. \qquad (S15)$$

It is worth mentioning that in this case, the input impedance and the reflection coefficient are independent of *x* used in the active tank. According the coupled-mode analysis, the circuits in Supplementary Figures 10a and 10c share the same eigenfrequencies. In the exact *PTX*-symmetric phase, applying the real eigenfrequencies to the input impedance in Eq. (S15) results in $Z_{in} = Z_0$ and thus zero reflection can be obtained in these frequencies. Supplementary Figure 10d shows the reflection spectra for the *PTX*-symmetric circuits in Supplementary Figure 10c, under different values of *y*; here $\gamma = 2.5$, $\mu = 2.5$, $\eta = 0.2$, and *x* is an arbitrary positive real number (because $Z_{in}$ and *Γ* are independent of *x*). It is clearly seen that the zero reflection takes place at the same frequencies observed in Supplementary Figure 10b, and the resonance linewidth can be tuned by varying the value of *y*.